\documentclass[journal,transmag]{IEEEtran}

\usepackage{graphicx}
\usepackage{amssymb,amsmath}
\usepackage{lipsum}
\usepackage{xcolor}

\usepackage{lmodern}

\usepackage[numbers,sort&compress]{natbib}
\usepackage[colorlinks=true,allcolors=blue,breaklinks=true]{hyperref} 
\begin{document}

\title{DFT calculations of magnetocrystalline anisotropy energy\\ with fixed spin moment}


\author{
	\IEEEauthorblockN{
Justyn Snarski-Adamski\IEEEauthorrefmark{1},
Joanna Marciniak\IEEEauthorrefmark{1,2},
Wojciech Marciniak\IEEEauthorrefmark{1,2,3}, \\
Justyna Rychły-Gruszecka\IEEEauthorrefmark{1}, and
Mirosław Werwiński\IEEEauthorrefmark{1}
}	
	\IEEEauthorblockA{\IEEEauthorrefmark{1}Institute of Molecular Physics, Polish Academy of Sciences, Smoluchowskiego 17, 60-179 Poznań, Poland}
	\IEEEauthorblockA{\IEEEauthorrefmark{2}Department of Physics and Astronomy, Uppsala University, P.O. Box 516, 751 20 Uppsala, Sweden}      
    \IEEEauthorblockA{\IEEEauthorrefmark{3}Faculty of Materials Engineering and Technical Physics, Poznan University of Technology,\\ Rychlewskiego 1, 61-131 Poznań, Poland}	
  
}

\IEEEtitleabstractindextext{%
\begin{abstract}
The development of new-generation permanent magnets is based on experimental efforts and innovative theoretical tools for modeling magnetic properties.
Magnetocrystalline anisotropy energy (MAE) \textendash{} one of the main intrinsic properties of permanent magnets \textendash{} 
can be calculated using density functional theory (DFT).
However, MAEs determined with different exchange-correlation potentials can vary  widely.
We show how these seemingly contradictory results can be reconciled using the fully relativistic fixed spin moment (FR-FSM) method.
This is because the equilibrium pairs [MAE, $\boldsymbol{m_s}$] calculated with different exchange-correlation potentials overlap with the MAE($\boldsymbol{m_s}$) curve determined from the FR-FSM method ($\boldsymbol{m_s}$ denotes the spin magnetic moment).
The FR-FSM method also enables the hypothetical maximum MAE value for a given material to be estimated.
In the case of magnetic alloys, MAE(FSM) analysis allows the optimal alloying additions to be determined in order to improve the MAE value.
The high independence of the MAE($\boldsymbol{m_s}$) works well for exchange-correlation functionals such as LDA and GGA. 
However, it is not a universal measure and varies with changes in the value of the U parameter when using LDA+U or GGA+U functionals.
Concluding, the framework we describe for MAE \textit{versus} FSM calculations can be a useful tool in the design of new permanent magnets.

\end{abstract}

}

\maketitle

\pagestyle{empty}
\thispagestyle{empty}

\IEEEpeerreviewmaketitle

\section{Introduction}

%
The development of a new generation of permanent magnets with limited rare earth content relies on experimental and theoretical efforts~\cite{gutfleisch_magnetic_2011, cui_current_2018,  skokov_heavy_2018}.
Calculations using density functional theory (DFT) allow the determination of several intrinsic properties of permanent magnets; including magnetic moment, Curie temperature, and magnetocrystalline anisotropy energy (MAE)~\cite{cedervall_influence_2018,vishina_fe2c-_2023}. 
This work focuses on the latter.
%
%
The calculated MAE value then allows the anisotropy field to be determined and the upper limits of the magnetic hardness, coercivity field, and energy product to be estimated.
High positive MAE values \textendash{} indicative of uniaxial magnetic anisotropy \textendash{} are often pursued in tetragonal, hexagonal, and orthorhombic systems.

MAE is usually calculated as the difference between the total energies associated with magnetization easy and hard axes.
The spin-orbit coupling (SOC), which is the origin of magnetocrystalline anisotropy and links the direction of magnetization to the crystal lattice, must be included in the calculations. 
Determining MAE also requires high numerical precision and the use of a robust exchange-correlation potential.
%
%
However, even when these criteria are satisfied, the MAE is typically evaluated in the ground state (at zero kelvins), whereas values at room temperature and above are often of greater interest.
This is particularly true because, for many materials, the dependence of MAE on temperature can be complex. 
For example, it may resemble a sine wave~\cite{edstrom_magnetic_2015}, which further complicates the theoretical search for permanent magnets.

%
Developed in the 1980s, the fixed spin moment (FSM) method performs self-consistent calculations at a fixed total spin magnetic moment~\cite{schwarz_itinerant_1984,moruzzi_ferromagnetic_1986}.
It is a standard method implemented in many DFT codes.
In its non-relativistic and scalar-relativistic versions, it is used to study metamagnetic transitions and magnetic instabilities~\cite{schwarz_itinerant_1984,moruzzi_ferromagnetic_1986}.
Its applications also include: the determination of Landau coefficients~\cite{kuzmin_landau-type_2008}, volume magnetostriction~\cite{moruzzi_ferromagnetic_1986}, and exchange between sublattices~\cite{kuzmin_spin_2004-1}.
It also allows for the numerical stabilization of magnetic states that would be unstable in regular self-consistent calculations~\cite{lee_evaluation_2012, lee_half_2008,sims_theoretical_2012}.

%
The situation looks different in the case of \textit{fully relativistic} fixed spin moment (FR-FSM) approach.
The first mention on this topic dates back to 2011 and can be found in doctoral thesis of Carsten Neise~\cite{neise_magnetic_2011}, who used an unofficial version of the FPLO code~\cite{koepernik_full-potential_1999,opahle_full-potential_1999,eschrig_chapter_2004} to determine the MAE of UNi$_2$ for a fixed experimental value of the magnetic moment.
The official FPLO version implementing the FR-FSM was released in 2014 (FPLO14), and has since been perhaps the single publicly available implementation of this exact method.
In 2023, a group from Uppsala University presented the results of MAE calculations using the FR-FSM performed with the RSPt code~\cite{herper_magnetic_2023, wills_full-potential_2010}, indicating both the possibility and need for implementation of the method in other DFT codes.
In addition to the application of FR-FSM to MAE calculations, it is also useful, for example, for determining metamagnetic states in heavy-fermion materials (which require a fully relativistic description) and for stabilizing antiferromagnetic states (magnetic moment fixed to zero) in materials containing heavy elements.

%
How does the FR-FSM work in the FPLO?
A term containing auxiliary magnetic field ($B_{\rm{FSM}}$) \textendash{} needed to set the fixed spin moment \textendash{} is added to the fully relativistic Dirac equation. 
For a given fixed spin moment value, the initial guess of the auxiliary field is assumed.
Then, the spin magnetic moment is calculated and compared with the target value.
Subsequently, a suitable $\delta B_{\rm{FSM}}$ amendment is calculated to minimize this difference.
$B_{\rm{FSM}}$ and spin moment are then iteratively converged until a desired accuracy of the fixed spin moment is reached.
This method is an example of the implementation of the Lagrange multiplier method.

%
A different approach to constraining the magnetic moment is implemented in the VASP code. 
While the FPLO allows only collinear magnetic configurations with a fixed total spin magnetic moment, in the non-collinear version of VASP, the direction and magnitude of the magnetic moment on individual atoms in the unit cell can be specified.
However, VASP does not allow for directly fixing the total spin magnetic moment in calculations with SOC.
Nevertheless, using the projector augmented wave (PAW) formalism, accounting for SOC, and appropriately setting the magnetic moments on individual atoms, it should be possible to calculate the dependence of MAE on the spin magnetic moment, analogous to FR-FSM calculations in FPLO. 
The method used in VASP is called constrained local magnetic moments (or sometimes penalty functional method)~\cite{ma_constrained_2015,chen_constraining_2023}.

\begin{figure*}[!t]
\centering
\hfill{}
\includegraphics[clip,height = 0.65\columnwidth]{fept_mae_vs_m.eps}
\hfill{}
\includegraphics[clip,height = 0.65\columnwidth]{cefe12_mae_vs_m.eps}
\hfill{}
\vspace{5mm}\\
\hfill{}
\includegraphics[clip,height = 0.65\columnwidth]{feni_mae_vs_m.eps}
\hfill{}
\includegraphics[clip,height = 0.65\columnwidth]{feb_mae_vs_m.eps}
\hfill{}
\vspace{5mm}\\
\hfill{}
\includegraphics[clip,height = 0.65\columnwidth]{fe2p_mae_vs_m.eps}
\hfill{}
\includegraphics[clip,height = 0.65\columnwidth]{mnbi_pbe_vs_pw92.plusU.MAE_vs_FSM.eps}
\hfill{}
\caption{\label{fig:MAE_FSM} 
Overlap between MAE(FSM) results for various exchange-correlation potentials.
Magnetocrystalline anisotropy energies of selected rare-earth-free magnets are denoted by red circles.
We used exchange-correlation potentials 
in the form of:
von Barth-Hedin (BH),
Perdew-Wang (PW92),
Perdew-Burke-Ernzerhof (PBE),
and with exchange only.
Magenta (LDA) and blue (GGA) plots show the MAE as a function of fixed spin moment.
Calculations were made with the FPLO18 code.
These data were first published in Refs.: 
FePt~\cite{marciniak_dft_2022};
CeFe$_{12}$~\cite{snarski-adamski_effect_2022};
FeNi~\cite{marciniak_magnetic_2024};
FeB~\cite{snarski-adamski_searching_2025};
Fe$_2$P~\cite{casadei_rare-earth-free_2025};
MnBi (in preparation).
}
\end{figure*}

\begin{figure*}[!t]
\centering
\hfill{}
\includegraphics[trim = 0 123 0 123,clip,width = 0.85\columnwidth]{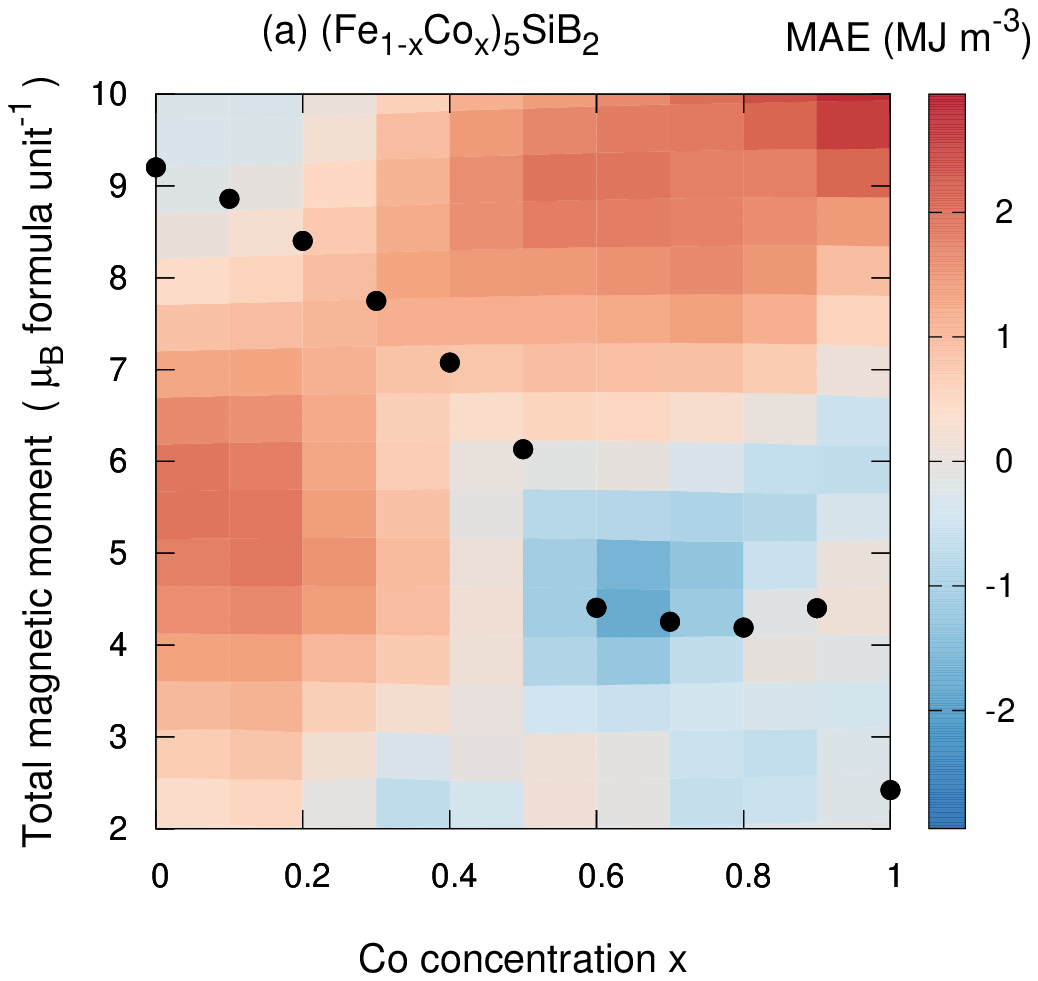}
\hfill{}
\includegraphics[trim = 0 123 0 123,clip,width = 0.85\columnwidth]{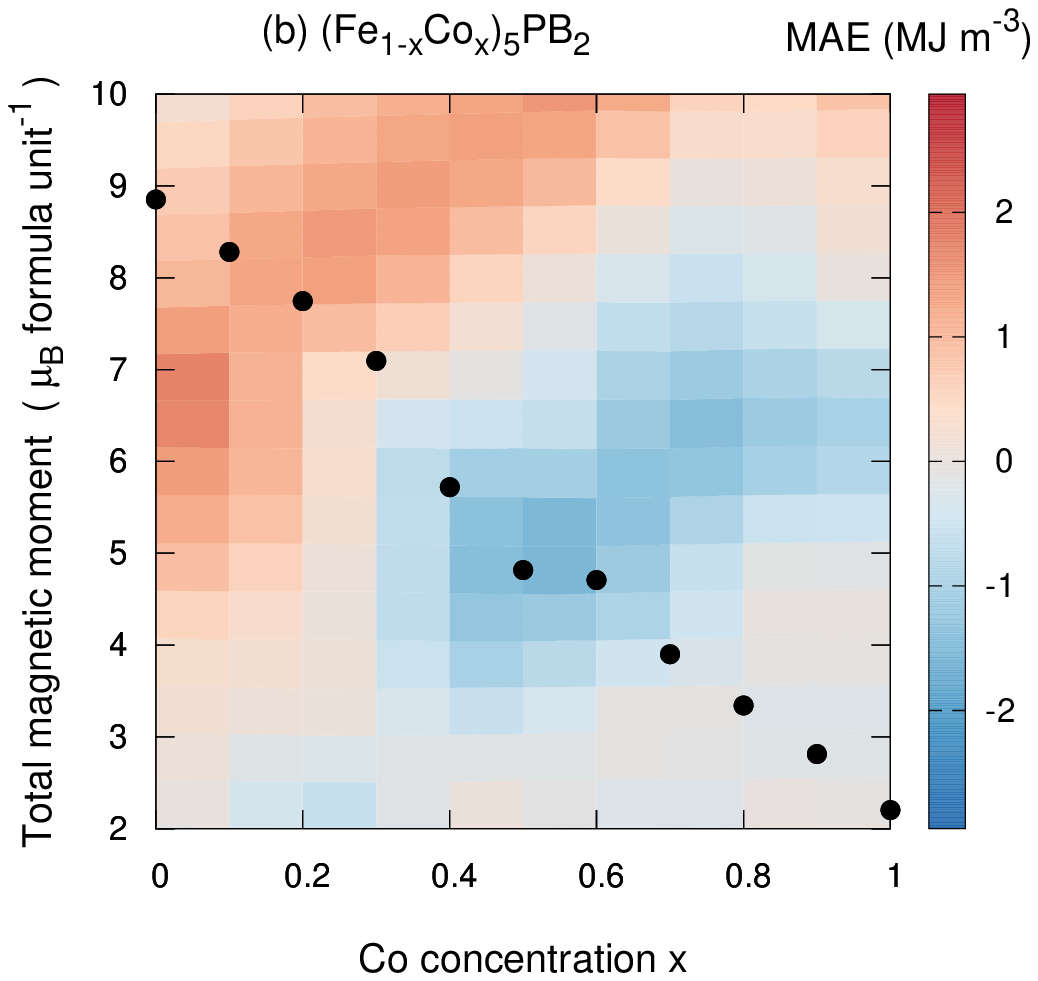}
\hfill{}
\caption{\label{fig:feco5sib2_feco5pb2} 
Magnetocrystalline anisotropy energies (MAE) of (a) (Fe$_{1-x}$Co$_{x}$)$_5$SiB$_2$  and (b) (Fe$_{1-x}$Co$_{x}$)$_5$PB$_2$ systems as a function of Co concentration $x$ and total magnetic moment.
Calculations were made with the FPLO14 code in the GGA-PBE approach.
The Co concentration was modeled using the virtual crystal approximation (VCA). 
We have set the values of the magnetic moments using the fixed spin moment (FSM) method.
The black circles indicate equilibrium solutions.
Data in panel (a) were first published in Ref.~\cite{werwinski_magnetic_2016}.
}
\end{figure*}

\section{Computational details}

%
MAE of a realistic, equilibrium system can be expressed based on contributions from the orbital magnetic moment and spin-resolved density. 
Such analysis can be exemplified by the detailed description of MAE in a tetragonal crystal by Miura and Okabayashi~\cite{miura_understanding_2022_1}. 
They combine an orbital magnetic-moment contribution to MAE, as described by Bruno~\cite{bruno_tight-binding_1989}, with a second term introduced by van der Laan~\cite{laan_microscopic_1998_1}. 
This second contribution, the intra-atomic magnetic dipole term, can be re-expressed in terms of orbital-projected spin moments. 
Conversely, artificially fixing the spin magnetic moment should then lead to a continuous, well-defined change in MAE. 
Many of the assumptions made in the derivation of this formula hold beyond the tetragonal crystal geometry. 
Hence, we expect MAE to form a continuous, non-trivial dependence on the spin magnetic moment.

%
To calculate the explicit MAE($m_s$) dependence, we used the usual energy difference method. We impose the ($m_s$) value by the fully relativistic fixed spin moment (FR-FSM) method in the FPLO code (versions 14 and 18)~\cite{koepernik_full-potential_1999,opahle_full-potential_1999,eschrig_chapter_2004}.
The MAE results presented in this work span several materials, and the calculation parameters (such as lattice parameters, number of $k$-points, and convergence criteria) have been reported previously in individual papers to which we refer further~\cite{marciniak_dft_2022,snarski-adamski_effect_2022,marciniak_magnetic_2024,snarski-adamski_searching_2025,casadei_rare-earth-free_2025}.
In order to determine the small magnitude of the MAE, we set up a very dense mesh of $k$-points (of the order of 30$^3$) and a high convergence criterion for the electron density (around 10$^{-7}$).
We also tested the MAE convergence each time.
When determining the MAE using the FR-FSM method, the computational load increases by a factor of approximately 50 compared to standard MAE calculations, as it is necessary to perform a series of independent calculations for a range of fixed magnetic moment values, as well as optimally apply both the LDA and GGA exchange-correlation functionals.
By default, calculations were performed using the generalized gradient approximation (GGA) in the Perdew-Burke-Ernzerhof (PBE) parameterization~\cite{perdew_generalized_1996}, which leads to relatively accurate magnetic moments for transition metals~\cite{tran_performance_2007}.
For comparison, we also performed MAE calculations using \textit{exchange-only} functional, and functionals from the family of local density approximation (LDA): 
von~Barth and Hedin (BH)~\cite{barth_local_1972},
Perdew and Zunger (PZ)~\cite{perdew_self-interaction_1981}, and
Perdew and Wang (PW92)~\cite{perdew_accurate_1992}.
Each time, we used the tetrahedron method to integrate over the Brillouin zone, as the Methfessel-Paxton method significantly affected the MAE results~\cite{blochl_improved_1994,methfessel_high-precision_1989}.
To determine the total energy of the system for a specific magnetization direction, we first performed self-consistent calculations in scalar-relativistic mode and then a single iteration in fully relativistic mode. 
This approach is fast and leads to minor differences in MAE compared to self-consistent fully relativistic calculations.
We presented a comparison of the results obtained from a single fully relativistic iteration and self-consistent calculations in Fig.~4 of Ref.~\cite{snarski-adamski_effect_2022}.

%
The method of determining MAE using FR-FSM has certain limitations. 
Calculating MAE requires using the full potential (or its very good approximation, e.g., PAW) and performing the calculations with a very dense $k$-point mesh, which makes the calculations computationally demanding.
In addition, we want to highlight potential issue that the MAE(FR-FSM) calculations may encounter.
We calculate the MAE as the difference between the energies determined for the hard and easy magnetization axes.
In equilibrium calculations, the spin magnetic moments in orthogonal directions may generally differ.
Whereas, in MAE(FSM) calculations, we assume exactly the same values of the total (fixed) spin magnetic moment for both magnetization directions considered.
While the systems studied here seems insensitive to this approximation, non-physical results may occur in cases where magnetic moments differ significantly between magnetization directions.
Typical equilibrium $|m_{s 100} - m_{s 001}|$ values are on the order of 0.001~$\mu_B$\,atom$^{-1}$ for 3$d$ metals (Fe, Co, Ni) and 0.01~$\mu_B$\,atom$^{-1}$ for intermetallic compounds with high magnetic anisotropy such as FePt or YCo$_5$.
Larger differences in magnetic moments are observed for compounds containing actinides, and in such cases the FR-FSM method may fail.
As mentioned, the above remarks apply to equilibrium cases.
However, what would the difference in magnetic moments look like for significantly increased or decreased values of spin magnetic moments?
Calculations based on volume changes, which have a major impact on the system’s spin magnetic moment, may shed light on this issue.
Based on our results for MnBi (not shown), a volume change in the range from 0.8 to 1.2 of equilibrium volume leads to changes in the spin magnetic moment in the range from 1.55 to 1.99~$\mu_B$\,atom$^{-1}$.
The difference $|m_{s 100} - m_{s 001}|$ associated with these magnetic moments reach a maximum value of 0.005~$\mu_B$\,atom$^{-1}$ (for the largest volume of 1.2~$V_0$).
This suggests that the FR FSM method may not produce large errors in the MAE values even relatively far from the equilibrium value of the magnetic moment.
However, it is advisable to estimate the error range in each case based on the aforementioned calculations of volume change.

\section{Results and discussion}

\begin{figure}[!t]
\centering
\includegraphics[trim = 0 0 0 0,clip,width = 0.9\columnwidth]{fe5sib2_fsm_calc_MAE_vs_total_M_16x_comparison_v2.eps}
\caption{\label{fig:fe5sib2_MAE_vs_volume} 
Magnetocrystalline anisotropy energy (MAE) of Fe$_5$SiB$_2$ as a function of total magnetic moment.
The results from the fixed spin moment (FSM) method are shown in blue.
The volume change ($V/V_0$) calculations are presented in red.
In the latter case, we first determined MAE and $m$ as functions of volume, and then plotted them as MAE($m$).
Green circle represents equilibrium case.
Calculations were made with the FPLO14 code in the GGA-PBE approach.
These data were first published in Ref.~\cite{werwinski_magnetic_2016}.
}
\end{figure}

%
Despite the importance of DFT calculations for the design of new permanent magnets, the described limitations often lead to results that are questionable, contradictory, and significantly different from measurements.
Even after taking all precautions to ensure reliable and accurate calculations of such a subtle measure as MAE, the results may still be difficult to interpret.
Therefore, we consider the FR-FSM approach to be a useful tool that provides a new perspective on data that would otherwise remain ambiguous.

%
First, we would like to discuss the impact of the choice of exchange-correlation potential on the determined values of spin magnetic moments.
In Fig.~\ref{fig:MAE_FSM}, we have compiled our results for several materials: FePt~\cite{marciniak_dft_2022},
CeFe$_{12}$~\cite{snarski-adamski_effect_2022}, FeNi~\cite{marciniak_magnetic_2024},
FeB~\cite{snarski-adamski_searching_2025},
Fe$_2$P~\cite{casadei_rare-earth-free_2025}, and
MnBi.
We can see that the equilibrium points (red circles) obtained for several LDA/GGA exchange-correlation potentials ($V_\mathit{xc}$) follow a certain pattern.
Namely, equilibrium magnetic moments increase with $V_\mathit{xc}$ in the following order: BH, PW92, PBE, \textit{exchange only}.
(The same trend we have also observed in Fe$_3$C and Co$_3$C~\cite{snarski-adamski_magnetic_2025}.)
This specific order results from the different treatment of electron correlations in individual functionals.
The differences in spin magnetic moment values obtained for various LDA/GGA functional are approximately a few percent.

The situation is different for MAE. 
For example, as can be seen in Fig.~\ref{fig:MAE_FSM}(b), the equilibrium MAE values for  CeFe$_{12}$ vary with change of $V_\mathit{xc}$ from approximately +2~MJ\,m$^{-3}$ (strong uniaxial anisotropy) to approximately $-1$~MJ\,m$^{-3}$ (in-plane anisotropy). 
Clearly, we are dealing with major qualitative change.
Incidentally, this example also well illustrates how doubts regarding the credibility of MAE calculations may arise in the eyes of experimenters.
However, the MAE values do not appear to be random. 
On the contrary, they overlap with the MAE(FSM) plot, which clearly indicates dependence of MAE on the spin magnetic moment.
This explains the discrepancies in MAE results across LDA/GGA.

%
The LDA- and GGA-based MAE(FSM) curves in Fig.~\ref{fig:MAE_FSM} also show considerable overlap, indicating a minor dependence of MAE(FSM) plots on the LDA/GGA choice. 
This suggests that MAE(FSM) curves are more general than MAE values alone.
Specifically, the MAE(FSM) dependence near equilibrium $m_S$ values can be viewed as a functional-independent relation, revealing a hypothetical maximum MAE value.

%
The MAE(FSM) analysis can be extended to alloys.
Figure~\ref{fig:feco5sib2_feco5pb2} shows the results for (Fe$_{1-x}$Co$_{x}$)$_5$SiB$_2$  and (Fe$_{1-x}$Co$_{x}$)$_5$PB$_2$ alloys in the virtual crystal approximation (VCA).
(
The conclusions based on the VCA calculations presented here are qualitative in nature, particularly with regard to the MAE.
VCA maintains the general MAE trends but overestimates the MAE values~\cite{edstrom_magnetic_2015}.)
MAE($x,m$) maps broaden the perspective of interpreting MAE results.
Comparing panels (a) and (b), we can see that replacing Si with P, which adds one electron per formula unit, shifts the MAE map toward the upper left corner.
These results allow us to propose further directions for optimizing these alloys.
For example, through alloying the 3$p$ element (Si/P)~\cite{hedlund_magnetic_2017} or additional alloying with selected transition metals aimed at changing the magnetic moment in a way that correlates with an increase in MAE~\cite{edstrom_magnetic_2015,werwinski_magnetic_2016,werwinski_magnetocrystalline_2018}.

%
{
Figure~\ref{fig:fe5sib2_MAE_vs_volume} shows the MAE(FSM) results for Fe$_5$SiB$_2$.
For this compound, we also calculated the MAE and magnetic moment as a function of volume change ($V/V_0$) within a range of $\pm$~8\% relative to the equilibrium volume ($V_0$).
As expected, the magnetic moment increases with increasing volume (not shown).
In Fig.~\ref{fig:fe5sib2_MAE_vs_volume} we present the results of volume change as the dependence of MAE on the corresponding magnetic moment.
The figure shows that the MAE($m_{\rm s}$) dependencies are similar regardless of whether the magnetic moment was varied with FSM method or volume change.
The observed similarity in the relationships suggests that the analysis of the MAE as a function of magnetic moment can be performed also in codes without an implemented FR-FSM method by varying the volume.
Furthermore, a detailed analysis of the impact of FSM and alloying on the band structure and the MAE is presented in Ref.~\cite{werwinski_magnetic_2016}.
In Ref.~\cite{werwinski_ab_2017}, we also described how MAE contributions manifest at the Fermi level.

%
In justified cases of systems with stronger correlations, going beyond LDA/GGA is a reliable way to improve the agreement between computational and experimental results.
As shown by the results for MnBi, see Fig.~\ref{fig:MAE_FSM}(f), adding the Hubbard term (+U)~\cite{czyzyk_local-density_1994}, which takes into account additional correlations (in this case on Mn~3$d$ orbitals), affects the course of the MAE(FSM) plot.
Increasing the U value also causes a significant rise in the equilibrium spin magnetic moment.
These results show that MAE(FSM) relationships determined from LDA/GGA are not universal and may be insufficient to describe a given material.
Hence, the conclusion is that the MAE(FSM) relationship is not universal, and in the cases presented, the strong similarity in MAE(FSM) results is limited to LDA/GGA-type functionals, whereas the addition of the Hubbard U parameter can significantly affect the result.

\section{Summary and conclusions}
The fully relativistic fixed spin moment (FR-FSM) method for evaluating magnetocrystalline anisotropy energy (MAE) provides key insights into the properties of permanent magnets. 
It clarifies the discrepancies in MAE results across various exchange-correlation potentials and helps explain seemingly contradictory findings.
This method is largely independent of potential choice, allowing MAE(FSM) calculations to establish a hypothetical MAE maximum for a given system. 
It thus helps assess material suitability for permanent magnet applications and guides composition modifications tailored to enhance MAE.
The application of LDA+U/GGA+U-type functionals for MnBi significantly affects the MAE(FSM) curve and demonstrates that MAE(FSM) is not a universal quantity, but depends on the value of the U parameter. 
However, the observed trends in the MAE(FSM) are largely independent of the choice of LDA/GGA-type exchange-correlation functional, and for this type of approximation, they can serve as an important point of reference.
The use of the FR-FSM method in MAE research improves the interpretation of calculation results, potentially contributing to the development of a new generation of permanent magnets.
However, the practical application of the FR-FSM method is restricted by the availability of supporting computational codes. 
Currently, only two codes, FPLO and RSPt, have been proven to enable FR-FSM calculations.
As such, similar implementations in other codes offer an interesting avenue for the future.
Researchers whose DFT code does not include the FR-FSM implementation can use the volume dependence as an intermediate variable to vary the magnetic moment and thus perform the MAE($m_s$) calculation.

\section*{Acknowledgements}
MW acknowledges the financial support of the National Science Center Poland under the decision DEC-2021/41/B/ST5/02894 (OPUS 21).
JM and WM acknowledge financial support from the Knut and Alice Wallenbergs' Foundation (grant no. 2022.0079).
We thank P.~Leśniak and D.~Depcik for compiling the scientific software and managing the computer cluster at the Institute of Molecular Physics, Polish Academy of Sciences.
We are grateful to J\'an Rusz and Jorge Luis Brise\~no-G\'omez for valuable comments on the manuscript.

\nocite{apsrev42Control}
\bibliography{mae_fsm_intermag_2026}

@CONTROL{apsrev42Control,
  author="08",
  title="0",
  pages="1",
  volume="0",
  year="1"
}

@article{barth_local_1972,
  title = {A Local Exchange-Correlation Potential for the Spin Polarized Case. i},
  author = {von Barth, U. and Hedin, L.},
  year = 1972,
  month = jul,
  journal = {J. Phys. C Solid State Phys.},
  volume = {5},
  number = {13},
  pages = {1629},
  issn = {0022-3719},
  doi = {10.1088/0022-3719/5/13/012},
  urldate = {2024-01-18},
  langid = {english}
}

@article{blochl_improved_1994,
  ids = {blochl_improved_1994-1},
  title = {Improved Tetrahedron Method for {{Brillouin-zone}} Integrations},
  author = {Bl{\"o}chl, Peter E. and Jepsen, Ove and Andersen, Ole Krogh},
  year = 1994,
  journal = {Phys. Rev. B},
  volume = {49},
  number = {23},
  pages = {16223},
  urldate = {2016-02-07}
}

@article{bruno_tight-binding_1989,
  title = {Tight-Binding Approach to the Orbital Magnetic Moment and Magnetocrystalline Anisotropy of Transition-Metal Monolayers},
  author = {Bruno, Patrick},
  year = 1989,
  journal = {Phys. Rev. B},
  volume = {39},
  number = {1},
  pages = {865},
  doi = {10.1103/PhysRevB.39.865},
  urldate = {2016-11-23}
}

@misc{casadei_rare-earth-free_2025,
  type = {{{SSRN Scholarly Paper}}},
  title = {Rare-Earth-Free {{Fe2P-based}} Compounds: Tuning and Modeling of Hard Magnetic Properties},
  shorttitle = {Rare-Earth-Free {{Fe2P-based}} Compounds},
  author = {Casadei, Matteo and Werwi{\'n}ski, Miros{\l}aw and Belli, Matteo and Cugini, Francesco and Trevisi, Giovanna and Fabbrici, Simone and Cabassi, Riccardo and Allodi, Giuseppe and Pagnoni, Riccardo and De Juli{\'a}n Fern{\'a}ndez, C{\'e}sar and Fourn{\'e}e, Vincent and Gallino, Isabella and Busch, Ralf and Sanna, Samuele and Albertini, Franca},
  year = 2025,
  month = dec,
  number = {5929404},
  eprint = {5929404},
  publisher = {Social Science Research Network},
  address = {Rochester, NY},
  doi = {10.2139/ssrn.5929404},
  urldate = {2026-03-04},
  archiveprefix = {Social Science Research Network},
  langid = {english}
}

@article{cedervall_influence_2018,
  title = {Influence of {{Cobalt Substitution}} on the {{Magnetic Properties}} of {{Fe}}{\textsubscript{5}}{{PB}}{\textsubscript{2}}},
  author = {Cedervall, Johan and Nonnet, Elise and Hedlund, Daniel and H{\"a}ggstr{\"o}m, Lennart and Ericsson, Tore and Werwi{\'n}ski, Miros{\l}aw and Edstr{\"o}m, Alexander and Rusz, J{\'a}n and Svedlindh, Peter and Gunnarsson, Klas and Sahlberg, Martin},
  year = 2018,
  month = jan,
  journal = {Inorg. Chem.},
  volume = {57},
  number = {2},
  pages = {777--784},
  issn = {0020-1669, 1520-510X},
  doi = {10.1021/acs.inorgchem.7b02663},
  urldate = {2018-01-18}
}

@article{chen_constraining_2023,
  title = {Constraining Spin Directions in Density Functional Theory Calculations by Imposing a Local Magnetic Field},
  author = {Chen, Yingwei and Yang, Yali and Xu, Changsong and Xiang, Hongjun},
  year = 2023,
  month = jun,
  journal = {Phys. Rev. B},
  volume = {107},
  number = {21},
  pages = {214439},
  issn = {2469-9950, 2469-9969},
  doi = {10.1103/PhysRevB.107.214439},
  urldate = {2026-03-02},
  langid = {english}
}

@article{cui_current_2018,
  title = {Current Progress and Future Challenges in Rare-Earth-Free Permanent Magnets},
  author = {Cui, Jun and Kramer, Matt and Zhou, Lin and Liu, Fei and Gabay, Alexander and Hadjipanayis, George and Balasubramanian, Balamurugan and Sellmyer, David},
  year = 2018,
  month = jul,
  journal = {Acta Mater.},
  volume = {158},
  pages = {118--137},
  issn = {1359-6454},
  doi = {10.1016/j.actamat.2018.07.049},
  urldate = {2018-07-29}
}

@article{czyzyk_local-density_1994,
  title = {Local-Density Functional and on-Site Correlations: The Electronic Structure of {{La}}{\textsubscript{2}}{{CuO}}{\textsubscript{4}} and {{LaCuO}}{\textsubscript{3}}},
  shorttitle = {Local-Density Functional and on-Site Correlations},
  author = {Czy{\.z}yk, M. T. and Sawatzky, G. A.},
  year = 1994,
  journal = {Phys. Rev. B},
  volume = {49},
  number = {20},
  pages = {14211},
  doi = {10.1103/PhysRevB.49.14211},
  urldate = {2016-02-07}
}

@article{edstrom_magnetic_2015,
  title = {Magnetic Properties of ({{Fe}}{\textsubscript{1-x}}{{Co}}{\textsubscript{x}}){\textsubscript{2}}{{B}} Alloys and the Effect of Doping by 5d Elements},
  author = {Edstr{\"o}m, A. and Werwi{\'n}ski, M. and Iu{\c s}an, D. and Rusz, J. and Eriksson, O. and Skokov, K. P. and Radulov, I. A. and Ener, S. and Kuz'min, M. D. and Hong, J. and Fries, M. and Karpenkov, D. {\relax Yu}. and Gutfleisch, O. and Toson, P. and Fidler, J.},
  year = 2015,
  month = nov,
  journal = {Phys. Rev. B},
  volume = {92},
  number = {17},
  pages = {174413},
  issn = {1098-0121, 1550-235X},
  doi = {10.1103/PhysRevB.92.174413},
  urldate = {2015-11-30}
}

@incollection{eschrig_chapter_2004,
  title = {Chapter 12 - {{Relativistic}} Solid State Calculations},
  booktitle = {Theoretical and {{Computational Chemistry}}},
  author = {Eschrig, H. and Richter, M. and Opahle, I.},
  year = 2004,
  volume = {14},
  pages = {723--776},
  doi = {10.1016/S1380-7323(04)80039-6},
  urldate = {2017-02-15}
}

@article{gutfleisch_magnetic_2011,
  title = {Magnetic {{Materials}} and {{Devices}} for the 21st {{Century}}: {{Stronger}}, {{Lighter}}, and {{More Energy Efficient}}},
  shorttitle = {Magnetic {{Materials}} and {{Devices}} for the 21st {{Century}}},
  author = {Gutfleisch, Oliver and Willard, Matthew A. and Br{\"u}ck, Ekkes and Chen, Christina H. and Sankar, S. G. and Liu, J. Ping},
  year = 2011,
  month = feb,
  journal = {Adv. Mater.},
  volume = {23},
  number = {7},
  pages = {821--842},
  issn = {09359648},
  doi = {10.1002/adma.201002180},
  urldate = {2016-03-04}
}

@article{hedlund_magnetic_2017,
  title = {Magnetic Properties of the {{Fe}}{\textsubscript{5}}{{SiB}}{\textsubscript{2}} - {{Fe}}{\textsubscript{5}}{{PB}}{\textsubscript{2}}  System},
  author = {Hedlund, Daniel and Cedervall, Johan and Edstr{\"o}m, Alexander and Werwi{\'n}ski, Miros{\l}aw and Kontos, Sofia and Eriksson, Olle and Rusz, J{\'a}n and Svedlindh, Peter and Sahlberg, Martin and Gunnarsson, Klas},
  year = 2017,
  month = sep,
  journal = {Phys. Rev. B},
  volume = {96},
  number = {9},
  pages = {094433},
  issn = {2469-9950, 2469-9969},
  doi = {10.1103/PhysRevB.96.094433},
  urldate = {2017-09-28}
}

@article{herper_magnetic_2023,
  title = {Magnetic Properties of {{NdFe}}{\textsubscript{11}}{{Ti}} and {{YFe}}{\textsubscript{11}}{{Ti}}, from Experiment and Theory},
  author = {Herper, Heike C. and Skokov, Konstantin P. and Ener, Semih and Thunstr{\"o}m, Patrik and Diop, L{\'e}opold V. B. and Gutfleisch, Oliver and Eriksson, Olle},
  year = 2023,
  month = jan,
  journal = {Acta Mater.},
  volume = {242},
  pages = {118473},
  issn = {1359-6454},
  doi = {10.1016/j.actamat.2022.118473},
  urldate = {2026-01-07}
}

@article{koepernik_full-potential_1999,
  ids = {koepernik_full-potential_1999-1},
  title = {Full-Potential Nonorthogonal Local-Orbital Minimum-Basis Band-Structure Scheme},
  author = {Koepernik, Klaus and Eschrig, Helmut},
  year = 1999,
  journal = {Phys. Rev. B},
  volume = {59},
  number = {3},
  pages = {1743--1757},
  publisher = {American Physical Society},
  doi = {10.1103/PhysRevB.59.1743},
  urldate = {2016-02-07}
}

@article{kuzmin_landau-type_2008,
  title = {Landau-Type Parametrization of the Equation of State of a Ferromagnet},
  author = {Kuz'min, M. D.},
  year = 2008,
  month = may,
  journal = {Phys. Rev. B},
  volume = {77},
  number = {18},
  pages = {184431},
  issn = {1098-0121, 1550-235X},
  doi = {10.1103/PhysRevB.77.184431},
  urldate = {2026-03-02},
  copyright = {http://link.aps.org/licenses/aps-default-license},
  langid = {english}
}

@article{kuzmin_spin_2004-1,
  title = {Spin Reorientation in High Magnetic Fields and the {{Co-Gd}} Exchange Field in {{GdCo}}{\textsubscript{5}}},
  author = {Kuz'min, M. D. and Skourski, Y. and Eckert, D. and Richter, M. and M{\"u}ller, K.-H. and Skokov, K. P. and Tereshina, I. S.},
  year = 2004,
  month = nov,
  journal = {Phys. Rev. B},
  volume = {70},
  number = {17},
  pages = {172412},
  issn = {1098-0121, 1550-235X},
  doi = {10.1103/PhysRevB.70.172412},
  urldate = {2026-03-02},
  copyright = {http://link.aps.org/licenses/aps-default-license},
  langid = {english}
}

@article{laan_microscopic_1998_1,
  title = {Microscopic Origin of Magnetocrystalline Anisotropy in Transition Metal Thin Films},
  author = {van der Laan, Gerrit},
  year = 1998,
  month = apr,
  journal = {J. Phys. Condens. Matter},
  volume = {10},
  number = {14},
  pages = {3239},
  issn = {0953-8984},
  doi = {10.1088/0953-8984/10/14/012},
  urldate = {2026-03-07},
  langid = {english}
}

@article{lee_evaluation_2012,
  title = {Evaluation of Half-Metallic Antiferromagnetism in {{A}}{\textsubscript{2}}{{CrFeO}}{\textsubscript{6}} ({{A}} = {{La}}, {{Sr}})},
  author = {Lee, Kwan-Woo and Ahn, Kyo-Hoon},
  year = 2012,
  month = jun,
  journal = {Phys. Rev. B},
  volume = {85},
  number = {22},
  pages = {224404},
  issn = {1098-0121, 1550-235X},
  doi = {10.1103/PhysRevB.85.224404},
  urldate = {2026-03-02},
  copyright = {http://link.aps.org/licenses/aps-default-license},
  langid = {english}
}

@article{lee_half_2008,
  title = {Half Semimetallic Antiferromagnetism in the {{Sr}}{\textsubscript{2}}{{CrTO}}{\textsubscript{6}} System ({{T}} = {{Os}}, {{Ru}})},
  author = {Lee, K.-W. and Pickett, W. E.},
  year = 2008,
  month = mar,
  journal = {Phys. Rev. B},
  volume = {77},
  number = {11},
  pages = {115101},
  issn = {1098-0121, 1550-235X},
  doi = {10.1103/PhysRevB.77.115101},
  urldate = {2026-03-02},
  copyright = {http://link.aps.org/licenses/aps-default-license},
  langid = {english}
}

@article{ma_constrained_2015,
  title = {Constrained Density Functional for Noncollinear Magnetism},
  author = {Ma, Pui-Wai and Dudarev, S. L.},
  year = 2015,
  month = feb,
  journal = {Phys. Rev. B},
  volume = {91},
  number = {5},
  pages = {054420},
  issn = {1098-0121, 1550-235X},
  doi = {10.1103/PhysRevB.91.054420},
  urldate = {2026-03-02},
  copyright = {http://link.aps.org/licenses/aps-default-license},
  langid = {english}
}

@article{marciniak_dft_2022,
  title = {{{DFT}} Calculation of Intrinsic Properties of Magnetically Hard Phase {{L1}}{\textsubscript{0}} {{FePt}}},
  author = {Marciniak, Joanna and Marciniak, Wojciech and Werwi{\'n}ski, Miros{\l}aw},
  year = 2022,
  month = aug,
  journal = {J. Magn. Magn. Mater.},
  volume = {556},
  pages = {169347},
  issn = {03048853},
  doi = {10.1016/j.jmmm.2022.169347},
  urldate = {2022-04-29},
  langid = {english}
}

@article{marciniak_magnetic_2024,
  title = {Magnetic Anisotropy of {{L1}}{\textsubscript{0}} {{FeNi}} (001), (010), and (111) Ultrathin Films: {{A}} First-Principles Study},
  shorttitle = {Magnetic Anisotropy of {{L10}}{$<$}math{$><$}msub Is="true"{$><$}mrow Is="true"{$><$}/Mrow{$><$}mrow Is="true"{$><$}mn Is="true"{$>$}0{$<$}/Mn{$><$}/Mrow{$><$}/Msub{$><$}/Math{$>$} {{FeNi}} (001), (010), and (111) Ultrathin Films},
  author = {Marciniak, Joanna and Werwi{\'n}ski, Miros{\l}aw},
  year = 2024,
  month = nov,
  journal = {J. Magn. Magn. Mater.},
  volume = {609},
  pages = {172455},
  issn = {0304-8853},
  doi = {10.1016/j.jmmm.2024.172455},
  urldate = {2024-08-28}
}

@article{methfessel_high-precision_1989,
  title = {High-Precision Sampling for {{Brillouin-zone}} Integration in Metals},
  author = {Methfessel, M. and Paxton, A. T.},
  year = 1989,
  month = aug,
  journal = {Phys. Rev. B},
  volume = {40},
  number = {6},
  pages = {3616--3621},
  issn = {0163-1829},
  doi = {10.1103/PhysRevB.40.3616},
  urldate = {2022-02-02},
  langid = {english}
}

@article{miura_understanding_2022_1,
  title = {Understanding Magnetocrystalline Anisotropy Based on Orbital and Quadrupole Moments},
  author = {Miura, Yoshio and Okabayashi, Jun},
  year = 2022,
  month = nov,
  journal = {J. Phys. Condens. Matter},
  volume = {34},
  number = {47},
  pages = {473001},
  issn = {0953-8984, 1361-648X},
  doi = {10.1088/1361-648X/ac943f},
  urldate = {2023-12-15},
  langid = {english}
}

@article{moruzzi_ferromagnetic_1986,
  title = {Ferromagnetic Phases of Bcc and Fcc {{Fe}}, {{Co}}, and {{Ni}}},
  author = {Moruzzi, V. L. and Marcus, P. M. and Schwarz, K. and Mohn, P.},
  year = 1986,
  journal = {Phys. Rev. B},
  volume = {34},
  number = {3},
  pages = {1784},
  urldate = {2016-12-23}
}

@phdthesis{neise_magnetic_2011,
  title = {Magnetic {{Properties Studied}} by {{Density Functional Calculations Including Orbital Polarisation Corrections}}},
  author = {Neise, Carsten},
  year = 2011,
  month = jul,
  address = {Dresden, Germany},
  langid = {english},
  school = {Technischen Universit\"at Dresden}
}

@article{opahle_full-potential_1999,
  title = {Full-Potential Band-Structure Calculation of Iron Pyrite},
  author = {Opahle, I. and Koepernik, K. and Eschrig, H.},
  year = 1999,
  journal = {Phys. Rev. B},
  volume = {60},
  number = {20},
  pages = {14035},
  publisher = {American Physical Society},
  doi = {10.1103/PhysRevB.60.14035},
  urldate = {2016-02-07}
}

@article{perdew_accurate_1992,
  title = {Accurate and Simple Analytic Representation of the Electron-Gas Correlation Energy},
  author = {Perdew, John P. and Wang, Yue},
  year = 1992,
  journal = {Phys. Rev. B},
  volume = {45},
  number = {23},
  pages = {13244--13249},
  doi = {10.1103/PhysRevB.45.13244},
  urldate = {2016-02-07}
}

@article{perdew_generalized_1996,
  ids = {perdew_generalized_1996-1},
  title = {Generalized Gradient Approximation Made Simple},
  author = {Perdew, John P. and Burke, Kieron and Ernzerhof, Matthias},
  year = 1996,
  journal = {Phys. Rev. Lett.},
  volume = {77},
  number = {18},
  pages = {3865--3868},
  publisher = {American Physical Society},
  doi = {10.1103/PhysRevLett.77.3865},
  urldate = {2016-02-07}
}

@article{perdew_self-interaction_1981,
  title = {Self-Interaction Correction to Density-Functional Approximations for Many-Electron Systems},
  author = {Perdew, J. P.},
  year = 1981,
  journal = {Phys. Rev. B},
  volume = {23},
  number = {10},
  pages = {5048--5079},
  doi = {10.1103/PhysRevB.23.5048}
}

@article{schwarz_itinerant_1984,
  title = {Itinerant Metamagnetism in {{YCo}}{\textsubscript{2}}},
  author = {Schwarz, K. and Mohn, P.},
  year = 1984,
  journal = {J. Phys. F Met. Phys.},
  volume = {14},
  number = {7},
  pages = {L129},
  issn = {0305-4608},
  doi = {10.1088/0305-4608/14/7/008},
  urldate = {2016-05-02}
}

@article{sims_theoretical_2012,
  title = {Theoretical Investigation into the Possibility of Very Large Moments in {{Fe}}{\textsubscript{16}}{{N}}{\textsubscript{2}}},
  author = {Sims, H. and Butler, W. H. and Richter, M. and Koepernik, K. and {\c S}a{\c s}{\i}o{\u g}lu, E. and Friedrich, C. and Bl{\"u}gel, S.},
  year = 2012,
  month = nov,
  journal = {Phys. Rev. B},
  volume = {86},
  number = {17},
  pages = {174422},
  issn = {1098-0121, 1550-235X},
  doi = {10.1103/PhysRevB.86.174422},
  urldate = {2026-03-02},
  copyright = {http://link.aps.org/licenses/aps-default-license},
  langid = {english}
}

@article{skokov_heavy_2018,
  title = {Heavy Rare Earth Free, Free Rare Earth and Rare Earth Free Magnets - {{Vision}} and Reality},
  author = {Skokov, K.P. and Gutfleisch, O.},
  year = 2018,
  month = feb,
  journal = {Scr. Mater.},
  volume = {154},
  pages = {289--294},
  issn = {13596462},
  doi = {10.1016/j.scriptamat.2018.01.032},
  urldate = {2018-04-04},
  langid = {english}
}

@article{snarski-adamski_effect_2022,
  title = {Effect of Transition Metal Doping on Magnetic Hardness of {{CeFe}}{\textsubscript{12}}-Based Compounds},
  author = {{Snarski-Adamski}, Justyn and Werwi{\'n}ski, Miros{\l}aw},
  year = 2022,
  month = jul,
  journal = {J. Magn. Magn. Mater.},
  volume = {554},
  pages = {169309},
  issn = {03048853},
  doi = {10.1016/j.jmmm.2022.169309},
  urldate = {2022-04-12},
  langid = {english}
}

@article{snarski-adamski_magnetic_2025,
  title = {Magnetic Hardness of Hexagonal and Orthorhombic {{Fe}}{\textsubscript{3}}{{C}}, {{Co}}{\textsubscript{3}}{{C}}, ({{Fe}}--{{Co}}){\textsubscript{3}}{{C}}, and Their Alloys with Boron, Nitrogen, and Transition Metals: {{A}} First-Principles Study},
  author = {{Snarski-Adamski}, Justyn and Werwi{\'n}ski, Miros{\l}aw and {Rych{\l}y-Gruszecka}, Justyna},
  year = 2025,
  month = feb,
  journal = {APL Mater.},
  volume = {13},
  number = {2},
  pages = {021117},
  issn = {2166-532X},
  doi = {10.1063/5.0243334},
  urldate = {2025-02-26}
}

@article{snarski-adamski_searching_2025,
  title = {Searching for Magnetically Hard Monoborides (and Finding a Few): {{A}} First-Principles Investigation},
  author = {{Snarski-Adamski}, Justyn and Werwi{\'n}ski, Miros{\l}aw},
  year = 2025,
  month = oct,
  journal = {J. Magn. Magn. Mater.},
  volume = {629},
  pages = {173204},
  issn = {0304-8853},
  doi = {10.1016/j.jmmm.2025.173204},
  urldate = {2025-07-17}
}

@article{tran_performance_2007,
  title = {Performance on Molecules, Surfaces, and Solids of the {{Wu-Cohen GGA}} Exchange-Correlation Energy Functional},
  author = {Tran, Fabien and Laskowski, Robert and Blaha, Peter and Schwarz, Karlheinz},
  year = 2007,
  month = mar,
  journal = {Phys. Rev. B},
  volume = {75},
  number = {11},
  pages = {115131},
  issn = {1098-0121, 1550-235X},
  doi = {10.1103/PhysRevB.75.115131},
  urldate = {2026-03-02},
  copyright = {http://link.aps.org/licenses/aps-default-license},
  langid = {english}
}

@article{vishina_fe2c-_2023,
  ids = {vishina_fe_2023-1},
  title = {Fe{\textsubscript{2}}{{C-}} and {{Mn}}{\textsubscript{2}}({{W}}/{{Mo}}){{B}}{\textsubscript{4}}-Based Rare-Earth-Free Permanent Magnets as a Result of the High-Throughput and Data-Mining Search},
  author = {Vishina, Alena and Eriksson, O. and Herper, H. C.},
  year = 2023,
  month = jan,
  journal = {Mater. Res. Lett.},
  volume = {11},
  number = {1},
  pages = {76--83},
  issn = {2166-3831},
  doi = {10.1080/21663831.2022.2117576},
  urldate = {2022-09-28},
  langid = {english}
}

@article{werwinski_ab_2017,
  title = {Ab Initio Study of Magnetocrystalline Anisotropy, Magnetostriction, and {{Fermi}} Surface of {{L1}}{\textsubscript{0}} {{FeNi}} (Tetrataenite)},
  author = {Werwi{\'n}ski, Miros{\l}aw and Marciniak, Wojciech},
  year = 2017,
  month = dec,
  journal = {J. Phys. Appl. Phys.},
  volume = {50},
  number = {49},
  pages = {495008},
  issn = {0022-3727, 1361-6463},
  doi = {10.1088/1361-6463/aa958a},
  urldate = {2018-01-19}
}

@article{werwinski_magnetic_2016,
  title = {Magnetic Properties of {{Fe}}{\textsubscript{5}}{{SiB}}{\textsubscript{2}} and Its Alloys with {{P}}, {{S}}, and {{Co}}},
  author = {Werwi{\'n}ski, Miros{\l}aw and Kontos, Sofia and Gunnarsson, Klas and Svedlindh, Peter and Cedervall, Johan and H{\"o}glin, Viktor and Sahlberg, Martin and Edstr{\"o}m, Alexander and Eriksson, Olle and Rusz, J{\'a}n},
  year = 2016,
  month = may,
  journal = {Phys. Rev. B},
  volume = {93},
  number = {17},
  pages = {174412},
  doi = {10.1103/PhysRevB.93.174412},
  urldate = {2016-05-17}
}

@article{werwinski_magnetocrystalline_2018,
  ids = {werwinski_magnetocrystalline_2018-1},
  title = {Magnetocrystalline Anisotropy of {{Fe}}{\textsubscript{5}}{{PB}}{\textsubscript{2}} and Its Alloys with {{Co}} and 5d Elements: {{A}} Combined First-Principles and Experimental Study},
  shorttitle = {Magnetocrystalline Anisotropy of {{Fe}} 5 {{PB}} 2 and Its Alloys with {{Co}} and 5 d Elements},
  author = {Werwi{\'n}ski, Miros{\l}aw and Edstr{\"o}m, Alexander and Rusz, J{\'a}n and Hedlund, Daniel and Gunnarsson, Klas and Svedlindh, Peter and Cedervall, Johan and Sahlberg, Martin},
  year = 2018,
  month = dec,
  journal = {Phys. Rev. B},
  volume = {98},
  number = {21},
  pages = {214431},
  publisher = {American Physical Society},
  issn = {2469-9950, 2469-9969},
  doi = {10.1103/PhysRevB.98.214431},
  urldate = {2018-12-19}
}

@book{wills_full-potential_2010,
  title = {Full-{{Potential Electronic Structure Method}}},
  shorttitle = {Full-{{Potential Electronic Structure Method}}},
  author = {Wills, John M. and Alouani, Mebarek and Andersson, Per and Delin, Anna and Eriksson, Olle and Grechnyev, Oleksiy},
  year = 2010,
  month = dec,
  publisher = {Springer Science \& Business Media},
  googlebooks = {Cg335Ks4xhwC},
  isbn = {978-3-642-15144-6},
  langid = {english}
}

\end{document}